\documentclass[twocolumn,aps,prl,superscriptaddress] {revtex4}
\usepackage{graphicx}

\begin{document}

\title{Effect of surface roughness and adsorbates on superlubricity}

\author{V.N. Samoilov}
\affiliation{IFF, FZ-J\"ulich, 52425 J\"ulich, Germany}
\affiliation{Physics Faculty, Moscow State University, 117234 Moscow, Russia}

\author{C. Yang}
\affiliation{IFF, FZ-J\"ulich, 52425 J\"ulich, Germany}

\author{U. Tartaglino}
\affiliation{IFF, FZ-J\"ulich, 52425 J\"ulich, Germany}
\affiliation{Democritos National Simulation Center, Via Beirut 2, 34014 Trieste, Italy}

\author{B.N.J. Persson}
\affiliation{IFF, FZ-J\"ulich, 52425 J\"ulich, Germany}

\begin{abstract}

We study the sliding of elastic solids in adhesive contact with flat and
rough interfaces. We consider the dependence of the
sliding friction on the elastic modulus of the solids.
For elastically hard solids with planar surfaces with incommensurate surface
structures we observe extremely low friction (superlubricity), which
very abruptly increases as the elastic modulus decreases. We show that
even a relatively small surface roughness or a low concentration of
adsorbates may completely kill the superlubricity state.  

This manuscript has been published in the book:
``Superlubricity'', edited by A. Erdemir and J.-M. Martin,
Elsevier, Amsterdam (2007).
The copyright of the final version of this paper has been tranferred to
Elsevier.

\end{abstract}
\maketitle

{\bf 1. Introduction}
\vskip 0.3cm

Friction between solid surfaces is a very important phenomenon for
biology and technology \cite{P0} and it is very common in nature.
Static friction always involves the coexistence of different
metastable configurations at microscopic level.
When one surface slides on the other at low speed, first there is a
loading phase during which the actual configuration stores elastic
energy. Then, when the stored energy is large enough, an instability
arises \cite{Aubry,SSC,Nozier}: the system jumps abruptly to 
another configuration and releases
elastic energy into irregular heat motion. The exact way of how the
energy is dissipated does not influence the sliding friction force,
provided that the dissipation is fast enough to happen before the next
sliding event.

There are many possible origins of elastic instabilities, e.g., they may
involve individual molecules or, more likely, groups of molecules or
``patches'' at the interface, which have been named stress
domains \cite{P11,P2,Caroli,Caroli1}. As a result the overall motion may
not exhibit any stick and slip behavior at macroscopic level, since the
local rearrangements can occur at different times in an incoherent
manner. Moreover, at least at zero temperature, 
the friction force does not vanish in the limit of
sliding speed $v\to 0$, but it tends to some finite value which depends
on the average energy stored during the loading events.

On the other hand, a system without instabilities cannot present a non-vanishing 
kinetic friction as $v\to 0$: if for any sliding distance there exist only one stable
configuration, the energy stored at the interface is a single-valued
function of the relative sliding distance. Thus when sliding occurs
slowly the process has to be reversible, i.e., with negligible friction.

If the structure of the two contacting surfaces do not match, the
formation of pinned states is hindered: when some groups of
atoms are in registry with the other surface, occupying a local energy minimum, some
other groups of atoms cannot adjust to the local energy-minimum configuration
without a deformation of the solids.
In this case there is a competition between two energies: the lateral
corrugation of the interaction potential between the walls, and the
elastic energy to deform the solid so that every surface patch adjusts
into a local minimum. If the latter prevails, the system is pinned and
static friction appears. Otherwise, if the solid is sufficiently stiff,
local domains do not show any instability and can overcome reversibly
the local barriers. The overall effect is a motion with no static
friction, since when some domains move uphill some other regions move
downhill, so that the total energy is constant.
This absence of instabilities due to a mismatch of the two surfaces' structures
has been named {\em superlubricity}
\cite{Japan}, although a more appropriate name would have been {\em structural
lubricity} \cite{Muser}.

It is well known that elastically hard solids tend to exhibit smaller sliding friction
than (elastically) soft materials \cite{Elisa}. One extreme example is diamond which under normal
circumstances exhibits very low kinetic friction coefficient, of the order of 0.01, 
when diamond is sliding on diamond.
This can be explained by the nearly absence of elastic instabilities because of the 
elastic hardness of the material. 

Recently, superlubricity has been observed during sliding of graphite on
graphite: in the
experiment described in Ref.~\cite{Frenken} a tungsten tip with a graphite
flake attached to it is slid on an atomically flat
graphite surface. When the flake is in registry with the substrate stick-slip motion and large
friction are observed. When the flake is rotated out of registry, the forces felt by the different atoms
start to cancel each other out, causing the friction force to nearly vanish, and the contact to
become superlubric.

Graphite and many other layered materials are excellent dry lubricants. The most likely reason for this is that
the solid walls of the sliding objects get coated by graphite flakes or layers with different orientation
so a large fraction of the graphite-graphite contacts will be in the superlubric state. This will 
lead to a strong reduction in the average friction. However, the coated solid walls are unlikely to
be perfectly flat and clean,
and it is important to address how surface roughness and adsorbates may influence the superlubric state.
Here we will show that even a relatively small surface roughness 
or low adsorbate coverage may kill the superlubric state.

\vskip 0.5cm
{\bf 2. Model}
\vskip 0.3cm

The results presented below have been obtained for an elastic flat block sliding on
a rigid substrate. 
We considered both flat and rough substrates.
The atoms in the bottom layer of the block form a simple square lattice
with lattice constant $a$. The lateral dimensions $L_x=N_xa$ and $L_y=N_ya$. 
For the block, $N_x=N_y=48$. Periodic boundary conditions are
applied in the $xy$ plane.  
We have used a recently developed multiscale molecular dynamics model,
where the block
extends in the vertical $z$-direction a similar 
distance as along the $x$-direction \cite{Yang}
(see also \cite{PRE_2004}).

The lateral size of the block is equal to that of the substrate,
but for the latter we use a different lattice constant $b \approx a/\phi$, where $\phi=(1+\sqrt{5})/2$
is the golden mean, in order to hinder the formation of commensurate
structures at the interface. For the substrate, $N_x=N_y=78$.
The mass of a block atom is 197 a.m.u.\ and
the lattice spacing of the block is $a=2.6~\mbox{\AA}$, so to get the same
atomic mass and density of gold. The lattice spacing of the substrate 
is $b=1.6~\mbox{\AA}$. 
We consider solid blocks with different Young's moduli 
from $E=0.2 \ {\rm GPa}$ up to $1000 \ {\rm GPa}$. The Poisson
ratio of the block is $\nu = 0.3$. 

\begin{figure}
  \includegraphics[width=0.40\textwidth]{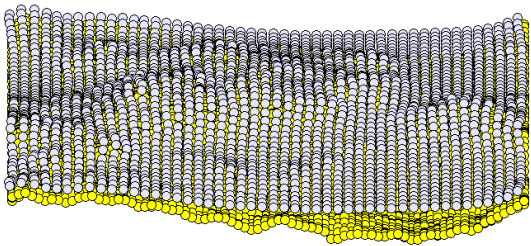} 
  \caption{ \label{contact1}
The contact between an elastic block with a flat surface and a rough 
{\it rigid} substrate. Only the interfacial layers of atoms are shown.
The elastic modulus of the block is $E=100 \ {\rm GPa}$. The substrate
is self-affine fractal with the root-mean-square roughness $3~\mbox{\AA}$,
fractal dimension $D_{\rm f} = 2.2$.
The applied pressure $p=10 \ {\rm 
GPa}$. 
}
\end{figure}

Many surfaces tend to be nearly self-affine fractal. A self-affine fractal
surface has the property that if part of the surface is magnified, with a magnification
which in general is appropriately different in the perpendicular direction to the surface as compared
to the lateral directions, then the surface ``looks the same'', i.e., the statistical
properties of the surface are invariant under this scale 
transformation \cite{Yang,P3}.
For a self-affine
surface the power spectrum has the power-law behavior
\begin{equation}
 \label{power_spectrum}
 C(q) \sim q^{-2(H+1)},
\end{equation}
where the Hurst exponent $H$ is related to the fractal dimension $D_{\rm f}$ of the surface via
$H=3-D_{\rm f}$. Of course, for real surfaces this relation only holds in some finite
wave vector region $q_0 < q < q_1$, and in a typical case $C(q)$ has 
a roll-off wavevector $q_0$ below which
$C(q)$ is approximately constant. 

In our calculations we have used self-affine fractal surfaces generated as outlined in 
Ref.~\cite{P3},
with the fractal dimension $D_{\rm f}=2.2$ and 
the roll-off wavevector $q_0=3q_L$, where $q_L= 2\pi /L_x$.
We have chosen $q_0=3q_L$ rather than $q_0=q_L$ since the former value gives 
some self-averaging and less noisy numerical results.
We also used $q_1= 2\pi /b = 78q_0$. 

The atoms at the interface between the block and the substrate interact with 
the Lennard-Jones potential
\begin{equation}
 \label{potential}
 U(r)=4 \epsilon \left [ \left ({r_0\over r}\right )^{12}-
 \left ({r_0 \over r}\right )^{6} \right],
\end{equation}
where $r$ is the distance between a pair of atoms.
In the calculations presented below we have used $r_0= 3.28~\mbox{\AA}$
and $\epsilon = 40 \ {\rm meV}$, which gives an 
interfacial binding energy (per unit area)
$\Delta \gamma \approx 4\epsilon/a^2 \approx 24 \ {\rm meV/\mbox{\AA}^2}$. 

As an illustration, in Fig.~\ref{contact1} we show
the contact between a flat elastic block (top) 
and a randomly rough 
{\it rigid} substrate (bottom). 
Only the interfacial block and substrate
atoms are shown. 
Note the elastic deformation of the block, and that non-contact regions
occur in the ``deep'' valleys of the substrate. 

In all results presented below 
the upper surface of the block 
moves with the velocity $v=0.1 \ {\rm m/s}$, and the (nominal) 
squeezing pressure $p$, if not stated otherwise, is
one tenth of the elastic modulus $E$ of the block, i.e., $p=0.1E$.
We did also some test calculations for $v=1 \ {\rm m/s}$ (not shown) 
but found very similar
results as for $v=0.1 \ {\rm m/s}$.
In fact, neglecting heating effects,
one does not expect any strong dependence of the friction force 
on the velocity,
as long as it is much smaller than the sound velocity (typically $1000 \ {\rm m/s}$), and
much higher than the velocities where thermally activated creep motion becomes
important (usually a few ${\rm \mu m/s}$). 
Furthermore, the sliding direction does not play a significant
role since the commensurability ratio 8/13 is the same along the $x$ and $y$ directions.
For the (randomly) rough surfaces we did some test calculations where we reversed the 
sliding direction and found that the friction force changed by at most 
20\%. This is a finite-size effect: for an infinite system sliding along the positive or negative
$x$-axis cannot change the friction force.

The reason for
choosing $p$ proportional to $E$ is twofold. First, we consider solids with elastic
modulus which varies over several orders of magnitude, and it is not possible to use 
a constant $p$
as this would result in unphysical large variations in the elastic deformation of the block.
Second, if two elastic solids are squeezed together with a given load, then as long as
the area of real contact is small compared to the nominal contact area, the 
pressure in the contact areas will be proportional to the elastic modulus of the
solids \cite{Persson_JCP2001}.

The system is kept at low temperature by 
a viscous friction 
(a Langevin thermostat at $T= 0 \ {\rm K}$)
applied only to the topmost layers of block atoms, which are far away from the
interface. 
On the time-scale of our simulations we observed 
no significant variation in the frictional shear force
which could be attributed to a (slow) heating up of the system.

\begin{figure}
  \includegraphics[width=0.42\textwidth]{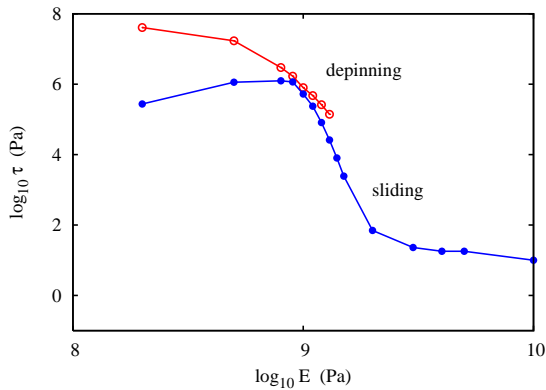} 
  \caption{ \label{stat_kin_Flat}
The shear stress at depinning (static friction) and during sliding
(kinetic friction) as a function of the elastic modulus 
$E$ of the block, for the flat substrate. 
In the calculation we have used the squeezing pressure $p = 0.1 E$ and
the sliding velocity $v=0.1 \ {\rm m/s}$.
}
\end{figure}

\begin{figure}
  \includegraphics[width=0.40\textwidth]{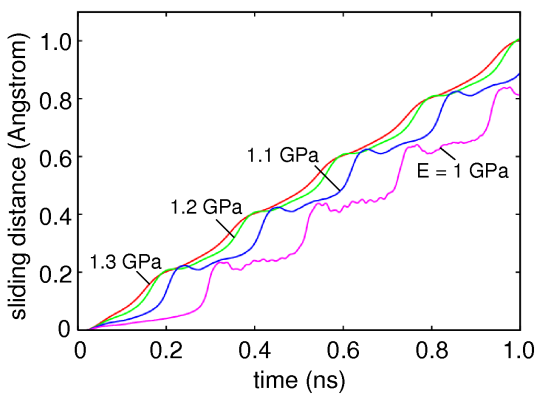} 
  \caption{ \label{sliding_distance_1}
The average displacement of the interface block atoms (in the sliding direction) 
as a function of time, for the flat 
substrate. The elastic modulus of the block is $E=1$, $1.1$, $1.2$ and $1.3 \ {\rm GPa}$.
The transition from high friction (stick-slip) to superlubricity (steady 
sliding) is demonstrated clearly.
}
\end{figure}

\vskip 0.5cm
{\bf 3. Numerical results}
\vskip 0.3cm

We first consider the sliding of clean smooth and rough surfaces for elastic solids
with different elastic modulus and surface roughness. We also study the influence of adsorbed
molecules on sliding friction. We show that already a small surface roughness or
less than a monolayer of adsorbed molecules may strongly increase the friction.

\vskip 0.3cm
{\bf 3.1. Clean smooth and rough surfaces}
\vskip 0.3cm

Let us first assume that both the block and the substrate have atomically smooth surfaces.
Fig.~\ref{stat_kin_Flat} shows the shear stress as a 
function of the elastic modulus $E$ of the block.
Note the relatively abrupt decrease in the friction when the elastic modulus changes
from $E_1\approx 0.7 \ {\rm GPa}$ to $E_2 \approx 2 \ {\rm GPa}$. 
For $E>E_2$ practically no instabilities occur and the
friction is extremely small,
while for $E < E_1$ relatively strong elastic
instabilities occur at the sliding interface, and the friction is high.
For $E=0.2 \ 
{\rm GPa}$ the static friction $\mu_{\rm s} > 2$. 
This calculation illustrates that the transition from high friction to {\it superlubricity} can 
be very abrupt; in the present case an increase in the elastic modulus by only a 
factor of 3 (from 0.7 to 2.1 GPa) 
decreases the kinetic friction by a factor of $\sim 10^{5}$.

We have studied the variation of
the shear stress as a function of time when the elastic modulus of the block
equals (a) $E= 0.8 \ {\rm GPa}$
and (b) $E= 2 \ {\rm GPa}$. 
When the elastic modulus of the solid is above the superlubricity threshold as in case (b), 
no significant elastic instabilities occur;
the stress is a periodic function of time,
with the period corresponding to the displacement 
$0.2 \ \mbox{\AA}$. For softer solids, when strong elastic instabilities 
occur during sliding as in case (a), the shear stress
is less regular (and the arrangement of the interfacial block atoms more disordered)
than for the stiffer solid, and the (average) period is {\it longer}
than $0.2 \ \mbox{\AA}$.  

The remarkable abruptness of the superlubricity transition is illustrated in
Fig. \ref{sliding_distance_1},
which shows the average displacement of the interface block atoms (in the sliding direction) 
as a function of time, for the flat 
substrate. Note the onset of stick-slip as 
the elastic modulus of the block decreases from $E=1.3 \ {\rm GPa}$ (upper curve) to
$1 \ {\rm GPa}$ (lower curve).

The regular pattern with period 0.2 \AA\ has a simple geometrical
explanation related to the commensurability ratio 8/13 along the sliding
direction: in the ground state each block's atom has 8 allowed
positions within the cell $b$ of the substrate. Hence there are 8
equivalent configurations within a sliding distance $b=1.6$ \AA.

\begin{figure}
  \includegraphics[width=0.42\textwidth]{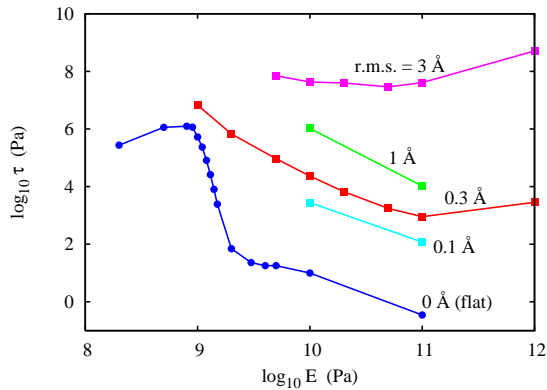} 
  \caption{ \label{KinFractal} 
The shear stress for an elastic block sliding on 
rough substrates, as a function of the elastic modulus $E$ of the block. 
The curves from top to bottom correspond to the root-mean-square roughness
amplitudes of the fractal substrate $3$, $1$, $0.3$, $0.1~\mbox{\AA}$ and $0$
(flat substrate). 
}
\end{figure}

Let us now consider the influence of surface roughness on the sliding dynamics.
In Fig.~\ref{KinFractal} we show the
average shear stress for an elastic block sliding on a
rough substrate, as a function of the elastic modulus $E$ of the block. 
For the substrate with the largest roughness, no superlubricity state can be observed
for any elastic modulus up to $E=10^{12} \ {\rm Pa}$.  

Our results are in agreement with the theoretical predictions of
Sokoloff \cite{Sokoloff}: roughness on multiple length scales can
induce a transition from a \emph{weak pinning regime} to a \emph{strong
pinning regime}. The main difference between our model and the one
of Sokoloff is that the latter assumes purely repulsive interactions between
the atoms, while here the attractive part of the Lennard-Jones potential
gives rise to an adhesive pressure $p_{\rm ad}$ 
which will contribute to the effective load.
Since in our case $p_{\rm ad}\approx 10$ GPa, the shear stress
that we obtain is almost independent of the external load.

\begin{figure}
  \includegraphics[width=0.40\textwidth]{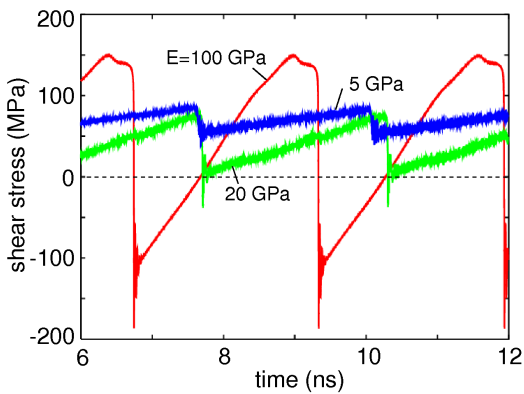} 
  \caption{ \label{shear1}
The shear stress as a function of time for the rough substrate
with the root-mean-square roughness amplitude $3~\mbox{\AA}$. The
elastic modulus of the block is $E=100$, $20$ 
and $5 \ {\rm GPa}$.
}
\end{figure}

\begin{figure}
  \includegraphics[width=0.40\textwidth]{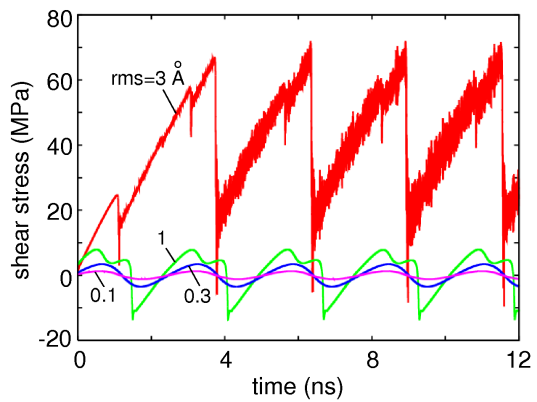} 
  \caption{ \label{shear2}
The shear stress as a function of time for the rough substrate with the 
root-mean-square roughness amplitudes $3$, $1$, $0.3$ and $0.1~\mbox{\AA}$. 
The elastic modulus of the block is $E=10 \ {\rm GPa}$. The transition 
from high friction to superlubricity is clearly demonstrated. 
}
\end{figure}

\begin{figure}
  \includegraphics[width=0.40\textwidth]{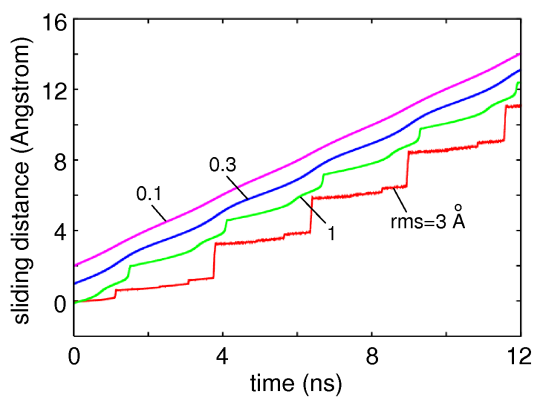} 
  \caption{ \label{sliding_distance}
The average displacement of the interface block atoms (in the sliding direction) 
as a function of time for the rough 
substrate with the root-mean-square roughness amplitudes $3$, $1$, $0.3$ and 
$0.1~\mbox{\AA}$. The elastic modulus of the block is $E=10 \ {\rm GPa}$.
The transition from high friction (stick-slip) to superlubricity (steady 
sliding) is clearly demonstrated.
}
\end{figure}

We have also studied the shear stress
as a function of time for different elastic moduli, 
see Fig. \ref{shear1}. 
Note that 
in addition to major slip events, several small slip events occur in all 
cases. These events correspond to local slip at some asperity contact regions
before the major slip involving the whole contact area.
In all cases, the time dependence of the shear stress remains periodic with the period  
$2.6~\mbox{\AA}$, which corresponds 
to the lattice spacing of the block.
At the current sliding speed $v=0.1$ m/s the kinetic friction force is smaller
for the stiffer solid even if the load is larger: although the maximum shear
stress achieved before sliding is quite big in such case, the average shear
stress is small and part of the elastic energy stored in the loading phase is
not dissipated, but it is given back to the system.
For the elastically softer
blocks ($E=20$ and $5 \ {\rm GPa}$), the stress-noise increases 
after each major slip event; this is caused by the elastic waves
(heat motion) excited during the (major) rapid slip events and not
completely adsorbed by the thermostat. 

In Fig. \ref{shear2} we show the effect of the substrate root-mean-square (rms) 
roughness amplitude on the shear stress as a function of time for the  block 
with the elastic modulus $E=10 \ {\rm GPa}$. We varied the root-mean-square 
roughness amplitude from $3$ to $0.1~\mbox{\AA}$. 
For the rms roughness amplitudes $0.3$ and $0.1~\mbox{\AA}$ the 
major slip is not as abrupt as for higher roughness amplitudes. In all 
cases, the time dependence of the shear stress remains periodic with the period 
$2.6~\mbox{\AA}$ determined by the lattice spacing of the block. For the 
rms roughness $3~\mbox{\AA}$ two small 
and a major slip events can be observed in each period, and the kinetic friction is 
high. For the case with the rms amplitudes $0.3$ and 
$0.1~\mbox{\AA}$ (almost) no elastic instability occurs, and the kinetic 
friction is very small.

In Fig. \ref{sliding_distance} we show 
the average displacement of the interface block atoms (in the sliding direction) 
as a function of time for the block with the elastic modulus $E=10 \ {\rm GPa}$, and 
for the rough substrate with various roughness amplitudes.  
The transition from high (stick-slip) friction for the most rough surface to 
very low friction (smooth sliding) for the surfaces with root-mean-square roughness
$0.3$ and $0.1 \ \mbox{\AA}$  
is demonstrated clearly.

\vskip 0.3cm
{\bf 3.2. Dependence of the friction on the load}
\vskip 0.3cm

We now study the dependence of the friction force on the load. We consider both flat surfaces
and the case where the substrate is rough. For flat surfaces the frictional shear stress
is independent of the squeezing pressure $p$ as long as $p$ is smaller than the adhesion
pressure $p_{\rm ad}$ which is of the order of $10^{10} \ {\rm Pa}$. For rough surfaces the 
situation
is more complex and the frictional shear stress will depend on the squeezing pressure $p$
even for very small $p$ unless at least one of the solids is so compliant that the attractive 
interaction becomes important for the contact mechanics.

\vskip 0.3cm
{\bf Flat surface}

\begin{figure}
  \includegraphics[width=0.40\textwidth]{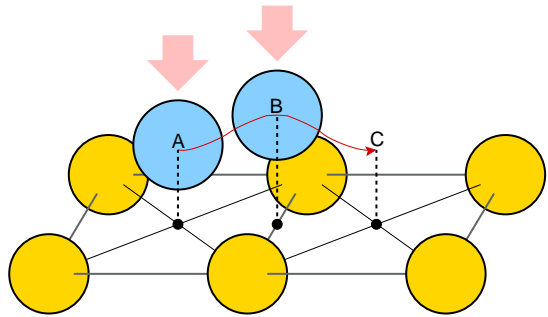} 
  \caption{\label{potential1}
A block atom moving (or jumping) from the hollow site {\bf A} over the bridge site {\bf B}
to the hollow site {\bf C}. The maximum energy position along the trajectory is at site {\bf B}.
}
\end{figure}

During sliding, the atoms at the sliding interface will experience energetic 
barriers derived from both the attractive interaction between the atoms
on the two opposing surfaces,
and from the applied load. Thus, we may define an {\it adhesion pressure}
$p_{\rm ad}$, and as long as
$p_{\rm ad} \gg p$, where $p$ is the pressure in the contact area derived from
the external load, the frictional shear stress will be nearly independent of the applied
load. Let us first consider
the limiting case where the elastic modulus of the block is extremely small. In this
case, in the initial pinned state (before sliding) all the block atoms will occupy
hollow sites on the substrate, as indicated by atom {\bf A} 
in Fig.~\ref{potential1}. During
sliding along the $x$-direction, the atom {\bf A} will move over the bridge
position {\bf B} and then ``fall down'' into the hollow position {\bf C} (we assume overdamped
motion). The minimum energy for this process is given by the barrier height 
$\delta \epsilon$
(the energy difference between the sites {\bf B} and {\bf A}) plus the work 
$pa^2 \delta h$
against the external load, where $a$ is the block lattice constant and $\delta h$ the change in the
height between sites {\bf B} and {\bf A} (which depends on $p$). Thus the frictional
shear stress $\sigma_{\rm f}$ is determined by $\sigma_{\rm f} a^2 b = \delta \epsilon+pa^2\delta h$, or
$$\sigma_{\rm f} = \delta \epsilon /(ba^2) +p\delta h/b = (p_{\rm ad}+p)\delta h /b,$$
where we have defined the adhesion pressure $p_{\rm ad} = \delta \epsilon/(a^2 \delta h)$.

In our case $\delta \epsilon \approx 3 \ {\rm meV}$ and $\delta h \approx 0.008 \ \mbox{\AA}$
giving $p_{\rm ad} \approx 10^{10} \ {\rm Pa}$. Thus, in the present case,
only when the local pressure in the
contact regions becomes of the order of $\sim 10 \ {\rm GPa}$, or more, it will start to
influence the shear stress. This result is in accordance with our simulation results.
Thus, for smooth surfaces, the shear stress acting on the block with the elastic modulus
$E=0.5 \ {\rm GPa}$, squeezed against the substrate with the pressure $p=50$ and
$150 \ {\rm MPa}$, is identical
($\approx 1 \ {\rm MPa}$) within the accuracy of the simulations.

\vskip 0.3cm
{\bf Rough surface}

In the absence of adhesion, contact mechanics theories predict that the area
of real contact $A$ between two elastic solids with randomly rough (but nominally flat)
surfaces is proportional to the squeezing force (or load) $F_{\rm N}$ as long as $A \ll A_0$
(where $A_0$ is the nominal contact area).
The law $A= \alpha F_{\rm N}$ holds strictly only if the roughness occurs on many different length
scales \cite{RevBP}, but this is (almost) always the case for natural surfaces (e.g., a stone surface) or 
surfaces of engineering interest. For an infinite system (thermodynamic limit) not only
$A$ is proportional to $F_{\rm N}$, but the 
distribution of sizes of the contact regions, and the stress distribution in the area of real 
contact, is independent of
the squeezing force if $A \ll A_0$. The physical picture behind these results
is that as the squeezing force increases, new contact regions
are formed in such a way that the quantities referred to above remain unchanged. However,
for any finite sized system this picture cannot hold exactly as it requires that some contact regions
have infinite-size, which is possible only for an infinite sized system. 

When the attractive interaction cannot be neglected, the area
of real contact is often assumed \cite{assumed} to be of the form  
$A \approx \alpha (F_{\rm N}+F_{\rm ad})$ where
$F_{\rm ad}$ represents an adhesive load, 
but this relation is only approximate \cite{RevBP,EUBP}. 
If the friction force $F_{\rm f}$
is assumed to be proportional to the area of real contact, then one expects
$F_{\rm f} =\sigma_{\rm c} A \approx \sigma_{\rm c} \alpha (F_{\rm N}+F_{\rm ad})$
so that the nominal frictional shear stress 
$$\sigma_{\rm f} = F_{\rm f}/A_0 \approx \sigma_{\rm c} \alpha (F_{\rm N}+F_{\rm ad})/A_0
= \sigma_{\rm c} \alpha (p+\sigma_{\rm a})\eqno(3)$$
where $p=F_{\rm N}/A_0$ is the (nominal) squeezing pressure and 
where the so called {\it detachment stress} \cite{RevBP,EUBP} $\sigma_{\rm a}$ gives 
a contribution
to the frictional stress from the attractive wall-wall interaction. From (3) it also follows that
the friction coefficient $\mu = \sigma_{\rm c} \alpha (1+\sigma_{\rm a}/p)$
will diverge as the squeezing pressure $p$ goes to zero.

\begin{figure}
  \includegraphics[width=0.40\textwidth]{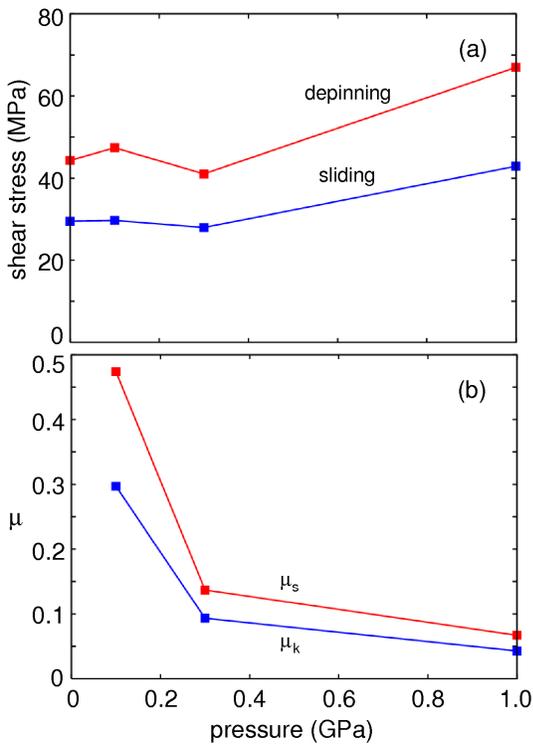} 
  \caption{ \label{av_shear_10}
The shear stress at depinning (static friction) and during sliding
(kinetic friction) (a) and the static and kinetic friction coefficients 
(b) as a function of the applied pressure, for an 
elastic block sliding on a rough substrate. For the substrate 
with the root-mean-square roughness amplitude $3~\mbox{\AA}$. The 
elastic modulus of the block is $E=10 \ {\rm GPa}$.
}
\end{figure}

We have studied the dependence of the frictional shear stress on the squeezing pressure
for the system studied above, for the block elasticities $E=10 \ {\rm GPa}$ and $E=100 \ {\rm GPa}$,
and for the rigid substrate with the rms roughness amplitude $3 \ \mbox{\AA}$.
In Fig. \ref{av_shear_10} we show (a) the (nominal) frictional shear stress and (b) the friction
coefficient as a function of the squeezing pressure when $E= 10 \ {\rm GPa}$. 
Note that, because of the attractive
interaction, the frictional shear stress
is constant for low squeezing pressures, while the friction coefficient diverges as
$p \rightarrow 0$.

\begin{figure}
  \includegraphics[width=0.40\textwidth]{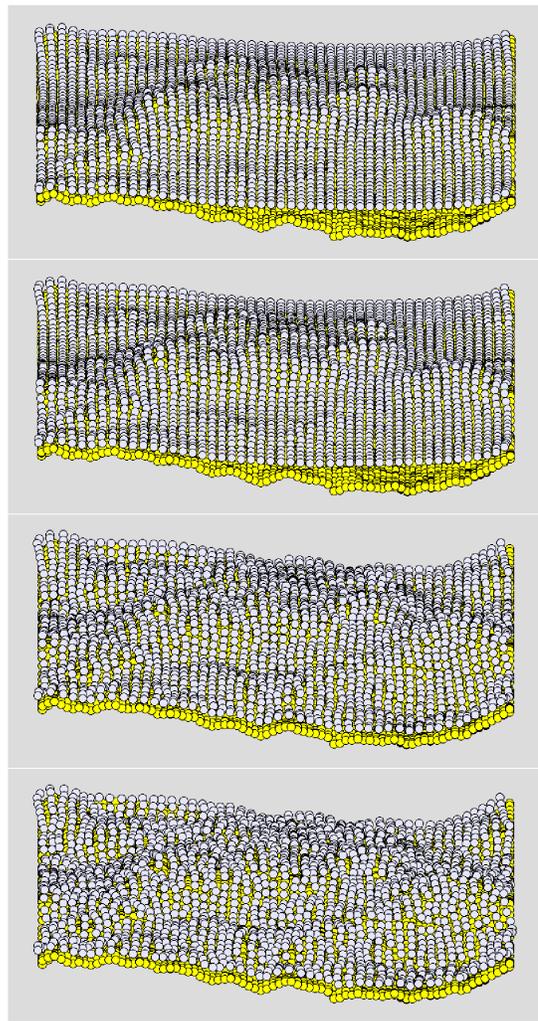} 
  \caption{ \label{contact2}
The contact between an elastic block with a flat surface and a rough 
{\it rigid} substrate. Only the interfacial layers of atoms are shown.
The elastic modulus of the block is $E=1000, 100, 10$ and $5 \ {\rm GPa}$ (from top to bottom). 
The substrate
is self-affine fractal with root-mean-square roughness $3~\mbox{\AA}$.
The applied pressure $p = 0.1 E$. 
Note the elastic deformation of the block, and that 
the real contact 
area is smaller than the nominal contact area for high values of 
the elastic modulus of the block. 
}
\end{figure}

The adhesion contribution $\sigma_{\rm c}\alpha\sigma_{\rm a}$
to the frictional shear stress becomes important only when the elastic modulus of the block
is small enough. The transition, with decreasing elastic modulus, 
from the case where the adhesion can be neglected to the case where it
is important, is rather abrupt. To illustrate this we show in Fig. \ref{contact2} the
interfacial atoms (the top atoms of the substrate and the bottom atoms of the block) for 
blocks with
the elastic modulus $E=1000$, 100, 10 and $5 \ {\rm GPa}$, and with the squeezing pressure
$p = 0.1 E$. In the absence of adhesion all the systems would
exhibit virtually identical arrangement of atoms. Indeed the two stiffest solids exhibit
very similar atom-arrangements, but there is a clear change when $E$ decreases 
from 100 to $10 \ {\rm GPa}$; in the latter case the attractive interaction is able to
pull the solids into intimate contact over most of the nominal contact area.
Thus the bottom surface of the block is able to bend and fill out a substrate
``cavity'' with diameter $D$ and height $h$ if the gain in wall-wall binding energy,
which is of the order of $D^2 \Delta \gamma$ (where $\Delta \gamma$ is the interfacial binding energy per unit
surface area for flat surfaces), is equal to (or larger than) the elastic energy stored in the deformation
field in the block, which is of the order of $ E D^3 (h/D)^2$. This gives the ``critical'' elasticity
$E_{\rm c} \approx D \Delta \gamma /h^2$. In the present case we have a distribution of roughness wavelength
but we can obtain a rough estimate of $E_{\rm c}$ by taking $h^2 = \langle h^2\rangle = 9 \ \mbox{\AA}^2$ as the mean of the
square of the roughness profile and 
$D \approx 100 \ \mbox{\AA}$ as a typical roughness wavelength.
Using $\Delta \gamma \approx 4 \epsilon /a^2 \approx 24 \ \mbox{meV/\AA}^2$ this gives
$E_{\rm c} \approx 40 \ {\rm GPa}$ which is between $100$ and $10 \ {\rm GPa}$. 
This change in the contact mechanics has a large influence on the sliding friction. Thus,
as we now will show, for the block with elastic modulus $E= 100 \ {\rm GPa}$ there will
be a negligible contribution to the friction from the attractive 
interaction and $\sigma_{\rm a} \approx 0$.

\begin{figure}
  \includegraphics[width=0.40\textwidth]{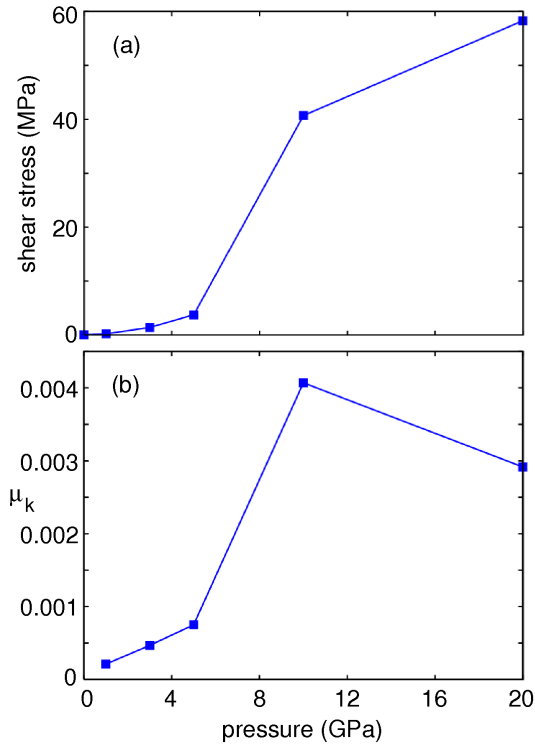} 
  \caption{ \label{av_shear_11}
The shear stress during sliding (kinetic friction) (a) and the kinetic 
friction coefficient (b) as a function of the applied pressure, for an 
elastic block sliding on a rough substrate. For the substrate 
with the root-mean-square roughness amplitude $3~\mbox{\AA}$. The 
elastic modulus of the block is $E=100 \ {\rm GPa}$.
}
\end{figure}

\begin{figure}
  \includegraphics[width=0.40\textwidth]{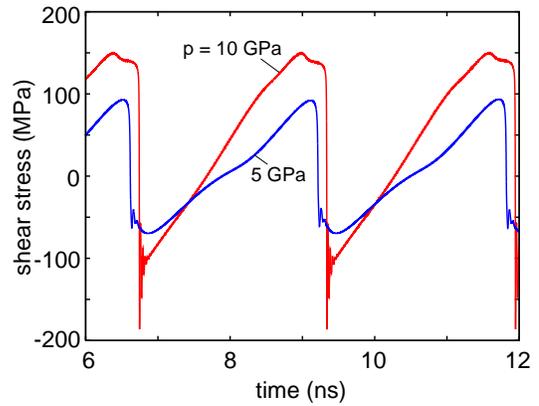} 
  \caption{ \label{Fig.5.10}
The shear stress as a function of time for the rough substrate
with the root-mean-square roughness amplitude $3~\mbox{\AA}$. The
elastic modulus of the block is $E=100  \ {\rm GPa}$
and the applied pressure $p=5$ and $10 \ {\rm GPa}$.
}
\end{figure}


Fig. \ref{av_shear_11} shows the frictional shear stress for 
the same system as in Fig. \ref{av_shear_10}
except that the elastic modulus of the block is ten times higher. In this case
the influence of the attractive interaction is {\it negligible}, and the frictional
shear stress decreases continuously as the squeezing pressure decreases. However,
the friction coefficient is not constant as expected from the arguments presented above 
related to the invariance of the pressure distribution and contact size distribution
with the squeezing force. This fact must be related to the small size of the system
used in our simulations. As the squeezing pressure increases the 
stress distribution at the interface and the average size of the contact regions
will change in such a way that when $p$ increases from
$p=5$ to $10 \ {\rm GPa}$ there is a very rapid increase in asperity contact regions
undergoing elastic instabilities during sliding.
This can be directly demonstrated by comparing the time-variation of the shear stress
for $p=5$ and $10 \ {\rm GPa}$, see Fig. \ref{Fig.5.10}.
Note that at the higher pressure some slip events take place before the main slip
event, i.e., new elastic instabilities appear and the frictional shear stress 
increases much faster than linear with the nominal squeezing pressure as $p$ increases from
$5$ to $10 \ {\rm GPa}$.

\begin{figure}
  \includegraphics[width=0.40\textwidth]{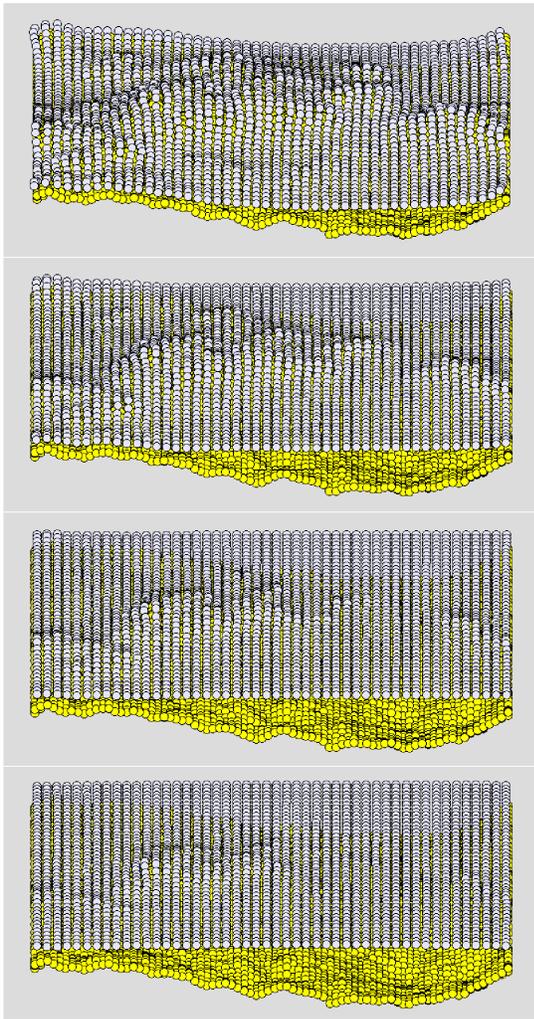} 
  \caption{ \label{contact3}
The contact between an elastic block with a flat surface and a rough 
{\it rigid} substrate. Only the interfacial layers of atoms are shown.
The elastic modulus of the block is $E=100 \ {\rm GPa}$. The substrate
is self-affine fractal with root-mean-square roughness $3~\mbox{\AA}$.
The applied pressure $p=10, 3, 1 \ {\rm 
GPa}$ and $0$ (from top to bottom). For the latter case only the 
adhesion pressure is acting. Note the elastic deformation of the 
block, and that the real contact 
area is smaller than the nominal contact area. 
}
\end{figure}

To illustrate the small influence of the adhesion on the contact mechanics for the
block with the elastic modulus $E=100 \ {\rm GPa}$ we show in Fig. \ref{contact3} 
the interfacial atoms for the squeezing pressures $p=10$, 3, $1 \ {\rm GPa}$ and 0.
When $p=0$ only the 
adhesion pressure is acting and the area of real contact almost vanishes.

\begin{figure}
  \includegraphics[width=0.42\textwidth]{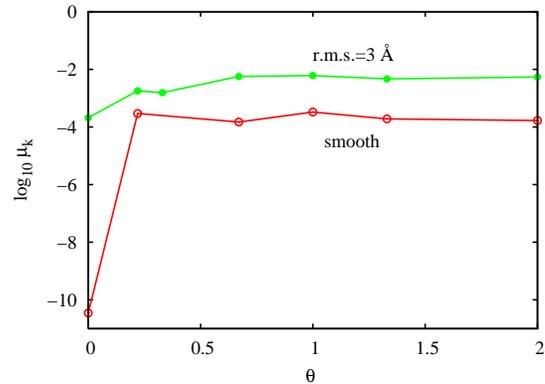} 
  \caption{ \label{adsorbate}          
          The kinetic friction coefficients for the elastic block 
          ($E=100 \ {\rm GPa}$)
          sliding both on smooth and on rough substrates, as a function of
          liquid coverage (octane, $\rm C_{8}H_{18}$) $\theta$ 
          confined between two walls \cite{Pulloff}. 
          The substrate has the root-mean-square
          (rms) roughness $3 \ \mbox{\AA}$ and the fractal dimension $\rm D_{f}=2.2$.
          The applied pressure $p = 1 \ {\rm GPa}$.
          } 
\end{figure}

\vskip 0.3cm
{\bf 3.3. Role of adsorbates}
\vskip 0.3cm

Extremely low sliding friction is possible only in the absence of elastic instabilities.
As shown above this is possible for stiff enough solids with incommensurate (or nearly incommensurate)
surface structures. However, any types of imperfections may ``lock'' the surfaces together
and introduce elastic instabilities during sliding. One type of ``defect'' discussed above
is surface roughness. Another possibility is adsorbed molecules. Adsorbed molecules may arrange themself
at the interface between the two solids in such a way as to pin the solids together.
A low concentration of (strongly bound) adsorbates is in many ways similar 
to nanoscale roughness and it is clear that if the perfect system 
(flat surfaces without adsorbates) is in a superlubric state,
one would expect a strong increase in the friction
already at low adsorbate coverage.
We have performed an extensive set of computer simulations to illustrate this
effect both for atomically smooth surfaces and for rough surfaces. 
In Fig. \ref{adsorbate} we show
the kinetic friction coefficients for the elastic block 
($E=100 \ {\rm GPa}$)
sliding both on smooth and on rough substrates, as a function of
liquid (octane, C$_{8}$H$_{18}$) coverage $\theta$ confined between two walls, 
for the applied pressure $p = 0.01 E$ \cite{Pulloff}. 
Note that for flat surfaces there is a very abrupt increase in the friction with
increasing adsorbate coverage. In fact, the friction increases by a factor of $\sim 10^{6}$
as the coverage increases from zero to $0.22$ monolayer. 
For $0.22 < \theta< 2$ the friction is nearly constant.
For the rough surface the increase in the friction is much smaller. In this case the (small)
increase in the friction results from octane molecules trapped in the asperity contact
regions \footnote{In the simulations the C$_8$H$_{18}$ bed-units interact with 
the solid wall atoms via the Lennard-Jones potential with the well-depth parameter
$\epsilon = 40 \ {\rm meV}$. This relatively strong interaction leads to 
lubricant molecules trapped in the asperity contact regions. Other studies 
(see Ref. \cite{Pulloff}) with $\epsilon = 5 \ {\rm meV}$ result in the 
squeeze-out of the lubricant from the asperity contact regions into the 
``valleys'' or ``cavities'' of the substrate height profile. In this case, 
which we will report on elsewhere \cite{Pulloff}, 
we do not expect any adsorbate-induced increase in the friction.}
-- this results in an effectively increased surface roughness and enhanced friction.

\vskip 0.5cm
{\bf 4. Summary and conclusion}
\vskip 0.3cm

To summarize, we have studied the sliding of elastic solids in adhesive contact with flat and
rough interfaces. We considered the dependence of the
sliding friction on the elastic modulus of the solids.
For elastically hard solids with planar surfaces with incommensurate surface
structures we observe extremely low friction (superlubricity), which
very abruptly increases as the elastic modulus decreases. 
Thus, at the superlubricity threshold,
an increase in the elastic modulus by a factor of $\sim 3$ resulted in
the decrease in the frictional shear stress by a factor $\sim 10^{5}$.
We have shown that
even a relatively small surface roughness, or a low concentration of adsorbates,
may completely 
kill the superlubricity.

\vskip 0.3cm
{\bf Acknowledgments}
\vskip 0.3cm

A part of the present work was carried out in frames of the ESF 
program ``Nanotribology (NATRIBO)''. Two of the authors (U.T. and 
V.N.S.) acknowledge support from IFF, FZ-J\"ulich, hospitality 
and help of the staff during their research visits.

\end{document}